\documentclass[runningheads]{llncs}
\usepackage[T1]{fontenc}
\usepackage{graphicx}
\usepackage{booktabs}
\usepackage{tabularx}
\usepackage[disable]{todonotes}
\usepackage{array}
\usepackage{amssymb}
\usepackage{csquotes}
\usepackage{marvosym}
\begin{document}
%

\title{Adopting InnerSource in a Public Sector Organization: A Case Study}
%
%

\author{Muhammad Usman\inst{1}\textsuperscript{(\Letter)} \orcidID{0000-0002-8132-0107} \and
Johan Linåker\inst{2} \orcidID{0000-0001-9851-1404} \and
Deepika Badampudi\inst{1} \orcidID{0000-0002-6215-1774} \and
Hannes Salin\inst{3} \orcidID{0000-0001-6327-3565}
}
\authorrunning{Usman et al.}
%

\institute{Department of Software Engineering, Blekinge Institute of Technology, Sweden.
\email{muhammad.usman, deepika.badampudi@bth.se}\\ \and
RISE Research Institutes of Sweden.
\email{johan.linaker@ri.se} \\ \and
Swedish Transport Administration, Sweden. \\
\email{hannes.salin@trafikverket.se}
}

\maketitle              
\begin{abstract}
InnerSource (IS) is increasingly viewed as a way to improve large-scale software development. However, evidence on its adoption in public sector organizations remains limited. This paper reports a study of IS adoption in a Swedish Public Sector Organization. Drawing on nine semi-structured interviews and an internal workshop with 13 participants, we examine motivations, challenges, and improvement strategies for IS adoption. The results show that the main motivations are improving reuse, strengthening cross-team collaboration, and enabling faster and more cost-efficient development. Key challenges relate to a lack of an open culture and mindset, management of reusable assets (e.g., ownership, documentation, and code visibility), and ways of working (e.g., siloed practices and limited time for contributions). The study also identified improvements to address challenges, including fostering an enabling culture and mindset for IS and knowledge sharing, strengthening management support, and establishing internal platforms and governance for reusable assets.

\end{abstract}
\section{Introduction}
\label{sec:Intro}
\todo[inline]{DEB}
Public sector organizations (PSOs) are increasingly developing and maintaining software-intensive systems to support their operations and service delivery. A census of Swedish government agencies illustrates this growth, with some agencies employing more than 100 developers \cite{borg2018digitalization}. In large organizations where software is developed and maintained across different units, challenges in coordination and reuse can arise. The census identified software reuse as a potential area for improvement \cite{borg2018digitalization}. Furthermore, as software development scales across organizations with multiple units and teams, it is crucial to coordinate and synchronize development efforts to improve reuse and avoid duplicate effort.

Innersource (IS) refers to the adoption of open source software development practices within organizational boundaries and has emerged as a promising solution to address the challenges of large-scale software development \cite{capraro2016inner,cooper2018adopting}. 
One of the key motivations is to decrease duplication of effort and improve internal reuse by making common reusable assets discoverable and easier to contribute to across organizational silos \cite{capraro2016inner,stol2014key,id20_chen2023using,id22_badampudi2025large}.
Previous studies also suggest that IS adoption is motivated by human and social factors, such as forming stronger internal communities, encouraging contributions, and improving employees' job satisfaction \cite{id19_wan2022motivates,stol2014key}.
While these benefits are well recognized, IS adoption and scaling do not come without challenges, such as resistance to change \cite{capraro2016inner}, a lack of management support \cite{id22_badampudi2025large}, insufficient time to contribute to IS projects, and concerns about ownership of IS projects \cite{state_of_innersource_2024,stol2014key,capraro2016inner}. Challenges such as these often require the establishment of structured governance, practices, and an overall cultural shift within the organization \cite{stol2014key,9758043,id20_chen2023using}.

Although the adoption and scaling of IS have been extensively studied in private-sector organizations, there is limited empirical evidence on IS adoption in PSOs. PSOs often have limited IT and engineering capabilities, due to a tradition of outsourcing and IT procurement~\cite{marco2020outsourcing,borg2018digitalization}. However, some PSOs have large internal IT divisions consisting of several sub-units and teams responsible for a large portfolio of software applications. The hierarchical structure of PSOs' IT divisions increases the likelihood of duplicate functionality, fragmented ownership of similar assets, and limited systematic reuse. Adopting IS is a potential solution to address some of these issues. However, there is not much research on IS adoption in PSOs (with some notable exceptions \cite{martin2011forge}). This study aims to address this gap by presenting a study of IS adoption in a PSO with a large IT division responsible for managing a large portfolio of applications.

Our study, through an exploratory case study at the Swedish Transport Administration (STA), represents an early effort to further explore this potential. We ask the following research questions (RQs):1) \textit{What is the motivation for adopting IS in PSOs?}, 2) \textit{What are the challenges in adopting IS in PSOs?}, and 3) \textit{How can the identified challenges be addressed?}
The findings provide empirical evidence on the motivations, challenges, and improvement strategies associated with IS adoption, offering a foundation for understanding how IS can be introduced and supported in PSOs.

\section{Research Methodology}
\label{sec:RM}
We conducted an exploratory case study \cite{runeson2012case} to answer the RQs.

\subsection{Case Organization and Unit of Analysis}

The case organization is the Swedish Transport Administration (STA), a Swedish PSO responsible for planning, construction, operation, and maintenance of Sweden's roads and railway infrastructure. STA operates in a large-scale, heavily regulated environment where digitalization and software systems play a central role, and has an annual budget of approximately 97 billion SEK (approximately 10.24 billion USD) \cite{trafikverket2025digitalization}.

Our study focuses in particular on the Information and Communication Technology (ICT) division at STA, which includes around 1,700 employees and contractors, with roughly 350–400 software developers and architects. Software development work is organized within a hybrid governance model combining the PM3 framework \cite{nordstrom2014pm3} with agile practices such as Scrum and Kanban.
 
An estimated 80-90 software engineering teams are located across four departments within the ICT division, either developing their own systems, customizing third-party systems, or maintaining commercial off-the-shelf (COTS) systems. In total, roughly 4-500 applications are developed. 

Within the context of the ICT division case at STA, the unit of analysis for our study is the development practices (e.g., development, maintenance, sharing, integration) related to common reusable assets. During the study period, no established IS practices were in place, although a smaller set of code repositories was available in the local TRV Hub. In addition, no official strategies or documents on open or inner sourcing were officially communicated within ICT at the beginning of the study.

\subsection{Data Collection and Analysis}
We used two data sources: 1) Data collected from nine semi-structured interviews for this study, and 2) Notes from an internal workshop at the case organization.

The relevant interviewees, representing a mix of roles and teams, were identified by the main contact person (fourth author) at STA (see Table \ref{tab:interviewees}. The interviews (half-hour long) were conducted remotely (via Microsoft Teams) and recorded with the permission of the interviewees. The fourth author did not participate in any interview. In addition to the questions about demographic information (e.g., role, experience, team), the interviews included questions about our three RQs. The interviewees were first asked about the motivation and need for adopting IS in the organization. Next, we asked them about the potential challenges of adopting IS practices and their suggestions for addressing them.

After the interviews, an internal workshop was held within the ICT division to discuss IS adoption. The workshop was attended by 13 participants (including the two interviewees) from the ICT division. The discussion points were recorded in a document, which we have used in this study as an additional data source.

\begin{table}
    \centering
    \scriptsize
    \begin{tabular}{p{3cm}p{9.5cm}}
    \toprule
        Role (STA experience) & Description \\
        \midrule
        1) Team lead (3 years)  & He is also a solution architect. He works with 15 people from different delivery sections in the transport and traffic control systems area.   \\
        2) Manager (7 years) & His two teams (15 people) work on the railroad and traffic systems. \\
        3) Team lead (2 years) & His team (eight members) works on railroad and traffic systems. \\
        4) Developer (1 year)  & His team (six members) is responsible for managing the developer platforms.\\ 
        5) Developer ( 8 years) & His team (two members) is responsible for the observability platform. \\
        6) Architect (4 years)  & He works with 8-10 teams that are responsible for case management, web, and e-commerce applications. \\
        7) Manager ( 6 years) & He manages two teams that are responsible for software solutions in technical documentation for the raidroads in Sweden. \\
        8) Developer (10 years) & His team (four members) is responsible for a system that manages how different contractors are doing infrastructure (e.g., roads). maintenance  \\
        9) Developer (6 years) & His team (ten members) is responsible for developing applications for internal users, mainly the business and management side.  \\
         \bottomrule
    \end{tabular}
    \caption{Interviewees}
    \label{tab:interviewees}
\end{table}

The Microsoft Team's auto-generated transcripts of the interviews were used for data analysis. The text segments were grouped and included under the broader topics that correspond to the aspects covered in the interview guide (i.e., motivation for IS, challenges, and improvement suggestions). The second and third authors then performed a thematic analysis \cite{braun2006using} to identify the themes under each of these topics. The identified themes were discussed and refined in a meeting (approximately one hour long in-person meeting) between the first three authors. Later, the analysis results were further discussed in a joint meeting (approximately 1.5 hours, onsite) among all four authors at one of the STA offices. The workshop document was also coded to identify the IS's motivating factors, challenges, and areas for improvement.

\section{Results}
\label{sec:Results}
This section presents the results of the RQs. 

\subsection{RQ1: Motivation for IS}

\begin{table}[h!]
\centering
\scriptsize
\setlength{\tabcolsep}{2pt}
\begin{tabularx}{\linewidth}{@{}c >{\raggedright\arraybackslash}X c c@{}}
\toprule
\textbf{No.} & \textbf{Factor} & \textbf{Interviews} & \textbf{Workshop} \\
\midrule

\multicolumn{4}{l}{\textbf{Motivators for IS adoption}} \\
\midrule
1 & To improve reuse across the entire organization & \checkmark & \checkmark \\
2 & To improve cross-team collaboration & \checkmark & \checkmark \\
3 & To achieve faster and cost-efficient development & \checkmark & \checkmark \\
4 & To lower maintenance effort & \checkmark & - \\
5 & To accelerate innovation & - & \checkmark \\
6 & To improve quality through more eyes & - & \checkmark \\
7 & To be an attractive and modern workplace & - & \checkmark \\

\midrule
\multicolumn{4}{l}{\textbf{Challenges in IS adoption}} \\
\midrule
1 & Lack of open mindset and culture & \checkmark & \checkmark \\
2 & Ownership of reusable assets & \checkmark & \checkmark \\
3 & Working in silos & \checkmark & \checkmark \\
4 & Code sharing and visibility & \checkmark & \checkmark \\
5 & Documentation of reusable assets & \checkmark & \checkmark \\
6 & Lack of time for reuse and collaboration & \checkmark & - \\
7 & Prioritizing own work & \checkmark & - \\
8 & Explaining InnerSource value & \checkmark & - \\
9 & Security clearance for code sharing & \checkmark & - \\

\midrule
\multicolumn{4}{l}{\textbf{Improvements to support IS adoption}} \\
\midrule
1 & Enabling culture and mindset for open collaboration & \checkmark & \checkmark \\
2 & Knowledge sharing across teams and units & \checkmark & \checkmark \\
3 & Internal platform for sharing reusable assets & \checkmark & \checkmark \\
4 & Documentation of reusable assets & \checkmark & \checkmark \\
5 & Upper management support and guidance & \checkmark & - \\
6 & Dedicated time for IS collaborations & \checkmark & - \\
7 & Meritocracy before hierarchy & - & \checkmark \\
8 & Clear responsibility and ownership & - & \checkmark \\
9 & Prioritize small changes and quick feedback & - & \checkmark \\
10 & Promote reuse and sharing of existing assets & - & \checkmark \\
11 & Clear processes and guidelines for contributions & - & \checkmark \\
\bottomrule
\end{tabularx}
\caption{Motivators, challenges, and improvements for InnerSource (IS) adoption.}
\label{tab:results}
\end{table}

Improving reuse is a leading motivator for adopting IS. Four interviewees highlighted it as the main reason behind IS adoption. It was also highlighted as a main factor in the workshop. An interviewee, while explaining the motivation for adopting the IS stated: \enquote{\textit{I think it's a good path to go. If it's reusable, then other people can use it and like extend it to their needs....You don't have to reinvent the wheel again. }}. Improving cross-team collaboration was also highlighted as a motivation for adopting IS, both in the interviews (two interviewees) and the workshop. In the workshop, IS is highlighted as a mechanism that could promote collaboration and help break down silos. An interviewee explained it as: \enquote{\textit{it's necessary that we cooperate more and to help each other with and use things from the code to do that. [IS] can be a way to make this happen, so we should cooperate more between the teams.}}. The interview (two interviewees) and workshop participants also highlighted faster and more cost-effective development as a motivating factor for adopting IS. This relates to reuse, i.e., IS helps improve reuse, which in turn helps teams developing with reuse to reduce their development costs and improve delivery time.

The workshop participants also identified additional motivating factors. IS could accelerate innovation within the organization by establishing mechanisms for improved cross-team collaboration, open and transparent processes, and knowledge-sharing opportunities. The workshop participants also highlighted that, since IS encourages code sharing, it would improve code quality by enabling code to be reviewed, reused, and contributed to by more people. Lastly, the workshop participants also reflected that adopting IS would help STA become an attractive and modern workplace where developers are encouraged and supported to work on interesting ideas and new technologies beyond their regular work, thereby helping them upskill their competence. Table \ref{tab:results} presents the list of motivating factors identified through interviews and the workshop.

\subsection{RQ2: IS adoption challenges}

Several common IS adoption challenges were identified in the interviews and the workshop (see Table \ref{tab:results} for details). First, the lack of an open mindset and culture for supporting cross-team and unit collaboration is identified as a challenge in the interviews (four interviewees) and the workshop. An interviewee highlighted it as: \enquote{\textit{It doesn't have anything to do with the governmental work or anything. It's just culture and mentality. And then some organizations do have a more the culture where that's more appreciated, which makes it easier.}} The ownership of reusable assets was also discussed as a potential challenge in the interviews (four interviewees) and workshop  - i.e., who will take the responsibility for the ownership and maintenance of the reusable asset. An interviewee explained it as: \enquote{\textit{If you take the initiative to start something and that you will be the the only one who maintains the code. That's I think it could be something that scares you off to start something like this.}}. Another challenge that acts as a barrier to IS adoption is that teams and units work in their silos, which limits the possibilities to identify and implement opportunities for co-developing common assets. Furthermore, the interview (four interviewees) and workshop participants also highlighted barriers in sharing and accessing code (e.g., due to security concerns) and other artifacts as a challenge in adopting IS in the organization. Finally, another common challenge in the interviews and the workshop relates to the documentation of reusable assets. For assets intended for reuse by multiple teams, the requirements for proper documentation are more stringent; otherwise, reusers may not be able to understand or reuse them.

\subsection{RQ3: Improvements}

Both interview and workshop participants shared several improvements to support IS adoption in the organization (see Table \ref{tab:results} for a complete list). First, the interview (three interviewees) and workshop participants suggested addressing the question of culture and mindset within the organization. They would like to see initiatives that foster a culture and mindset where collaboration across teams and units comes naturally. This, perhaps, also relates to Improvement 5 in Table \ref{tab:results}, which is about upper management support and guidance. An interviewee explained it as: \enquote{\textit{it's a mindset shift maybe that's need to take place that we should collaborate more because I think at the moment we are mostly
Looking inside each team and we solve everything inside the teams}}. Another interviewee reflected on it as: \enquote{\textit{The system developers need to talk to each other, not the managers or others. So it's a culture question, I think.}} Another related improvement suggestion is about creating opportunities for sharing knowledge across teams. The workshop participants suggested measures such as establishing chat channels, organizing hackathons, and teams organizing demos of their applications for others. Establishing an internal platform where the information about the IS projects and reusable assets (e.g., links to code repos, relevant owners, and related wikis and communication channels) is available for everyone in the organization was also shared as an improvement by interview (three interviewees) and workshop participants. This internal platform should allow efficient search and classification features so that it is easy to find relevant assets. The interview and workshop participants also suggested to improve the documentation of reusable assets (e.g., by AI-generated documentation).

The workshop participants also suggested some additional improvements. They suggested to promote meritocracy in the IS process so that the contributions and ideas are evaluated on the basis of their quality, not by the designation of the proposer. Potential contributors do not contribute to avoid becoming maintainers and owners of common assets. To address that, the workshop participants suggested clarifying the rules regarding ownership and appointing owners. The workshop participants also suggested that promoting reuse (over reinventing the wheel) should be an organizational focus area. Lastly, they also suggested formulating contributions guidelines so that contributors know what is expected from them when they send their commits and pull requests to a common asset.

\section{Discussion and Conclusion}
\label{sec:Discussion}

The findings highlight reuse as a main incentive, which, in many ways, can also be considered an enabler of other incentives, such as faster, more cost-efficient development. Higher degrees of collaboration and more eyes on the code may help accelerate innovation and quality, and decrease maintenance efforts, again leading to potentially faster and more cost-efficient development. In this sense, increased reuse or collaboration is not the end goal. The rationale for adopting IS should ideally be viewed holistically, considering cost, innovation, competency, culture, and employer branding as the main goals. These values can resonate with management and secure their support for introducing IS into the organization. In general, the incentives (improving reuse, fostering cross-team collaboration, and achieving faster and more efficient development and maintenance) align well with software-intensive organizations~\cite{id19_wan2022motivates,stol2014key}.

For challenges, a similar alignment between industry and the case organization is reiterated, e.g., regarding asset ownership, culture and mindset, siloed work practices, and incentivizing the adoption and implementation of IS~\cite{stol2014key,9758043,id20_chen2023using,edison2020inner}. To some degree, security clearance stands out, as some interviewees reported that their specific products were considered too sensitive or classified, even for internal disclosure. In STA's case, this can be seen in light of their responsibility for the physical infrastructure used for transport and mobility, as well as their role in national security. In terms of improvements, our observations reiterate industry best practices, e.g., establishing internal forges, better documentation, management support, dedicated time and mandate for collaboration, and advocacy and training for reuse~\cite{stol2014key,9758043,id20_chen2023using,id22_badampudi2025large,edison2020inner}.

The similarities, both in terms of perceived benefits, challenges, and improvement opportunities, indicate the potential to transfer practice from industry to a public-sector context. However, it should be noted that this should not serve as a potential generalization for PSOs at large. The STA has strong development capacity, with a complex team and product structure, which, anecdotally, can also translate to industrial contexts. PSOs, however, often have a tradition of outsourcing and procuring developer capacity, or solutions, with IT departments generally being small or limited to a product owner level~\cite{marco2020outsourcing,borg2018digitalization}. As we are limited by a single case, future research should investigate the potential for IS across PSOs more broadly with different types of characteristics. 

An aspect missing from our research and warranting special attention in future work is how the concept of IS can be extended to an inter-PSO (or intra-governmental) practice. PSOs within the same government may overlap and have similar needs to varying degrees, and internal sharing and reuse can be a means of transferring the same incentives for IS adoption from a single PSO to a cross-government level. The interviewees highlighted design systems for internal and external e-platforms as examples of solutions that are commonly redeveloped but where collaboration is now emerging, e.g., at the Swedish Public Employment Service. While the latter case concerns an open-source solution, there will be cases where a solution is not suitable for release as open source, yet can still be shared as a government-internal asset. Aspects such as security and classified access can be expected to prevent some sharing, as well as the generalizability of solutions and components. Still, in areas such as infrastructure and tooling, standardization and overlap may potentially be higher, as also indicated by the internal teams surveyed through the interviewees at STA.

The limitations of the study include a relatively small sample of interviewees in a large organization. To some extent, it was addressed by incorporating the workshop discussion into the analysis. The results were discussed jointly by all authors to avoid misinterpretation. In the future, we plan to expand the study by organizing workshops and focus groups at STA, to, for example, complement and prioritize challenges and improvements, and identify future focus areas.

\bibliographystyle{splncs04}
\bibliography{references}
%
\end{document}